\documentclass[12pt]{iopart}

\usepackage{iopams}
\usepackage{graphicx}
\usepackage{amssymb}
\usepackage{rotating}
\usepackage{pdflscape}
\usepackage{tablefootnote}
\usepackage{dcolumn}
\usepackage{bm}
\usepackage{cite}
\begin{document}

\title[]{Deformation dependence of 2p-radioactivity half-lives: Probe with a new formula across the mass region with Z$<$82}
\author{G. Saxena$^{1,\S}$, Mamta Aggarwal$^{2,\dagger}$, D. Singh$^{3}$, A. Jain$^{1,4}$, P. K. Sharma$^{5}$ and H. L. Yadav$^{6}$}
\address{$^{1}$Department of Physics (H\&S), Govt. Women Engineering College, Ajmer-305002, India}
\address{$^{2}$Department of Physics, University of Mumbai, Kalina, Mumbai-400098, India}
\address{$^{3}$Department of Physics, University of Rajasthan, Jaipur-302002, India}
\address{$^{4}$Department of Physics, Manipal University Jaipur, Jaipur-303007, India}
\address{$^{5}$Govt. Polytechnic College, Rajsamand-313324, India}
\address{$^{6}$Physics Department, Banaras Hindu University, Varanasi-221005, India.}

\ead{$^{\dagger}$mamta.a4@gmail.com, $^{\S}$gauravphy@gmail.com}
\vspace{10pt}
\begin{indented}
\item[]July 2022
\end{indented}

\begin{abstract}
Effect of deformation on half-life of two-proton (2p) radioactivity is investigated across the periodic chart for nuclei with Z$<$82. 2p-decay half-lives are estimated by employing our newly proposed semi-empirical formula wherein the nuclear deformation has been incorporated in a phenomenological way. Robustness of the formula is demonstrated as it estimates the measured values quite accurately and, hence, reliably applied to predict the other possible 2p-emitters. For many proton rich nuclei for which experimental data on the decay energies are not available, we have used the theoretical values obtained from our calculations using the relativistic mean-field (RMF) approach. The uncertainties in the theoretical decay energy values are minimised by machine learning (ML) technique. Correlation of 2p-radioactivity with 2p-halo and deformation is probed. Our calculations show the phenomenon of shape coexistence in several 2p-emitters, wherein the prolate shape is found to be more predominant for the ground state.\par
\end{abstract}

\submitto{\JPG}

\section{Introduction}
Two-proton radioactivity is a rare and lesser known exotic charged-particle decay mode, in which the valence protons are no longer bound by the strong nuclear forces and where the Coulomb and centrifugal barriers along with the structural effects prevent the proton emission and slow down the decay process that enables the life-times to be long enough to probe the atomic structure \cite{goldansky,WOODS,BLANKPLO}. The prediction of two-proton emission half-lives is generally difficult for the unknown nuclei as they crucially depend upon the accurate estimate of 2p-decay energy (Q$_{2p}$) evaluated with  high precision which itself is quite a challenging task in exotic nuclei. The deformation effects that increase the number of particles in the classically forbidden region below the continuum threshold and also contribute to pairing, influence the decay modes and the decay energy as well. Some recent works have speculated the phenomenon of shape coexistence, arising out of the competing shapes, to bring about the structural changes in the exotic nuclei and to affect the half-lives ~\cite{CRIDER,SARRIGUREN}, which has significant implications in the phenomena like r-process in nuclear astrophysics. In view of this possible significance of deformation in the stability of nuclear systems \cite{VM,MAPRC89}, Q$_{2p}$ and consequently the life-times, the structural effects must be incorporated in the evaluation of accurate Q$_{2p}$ and half-life values. The nuclei with Q$_{2p}$ $>$ 0 and Q$_p$ $<$ 0 define 2p-radioactivity \cite{Ascher,mukha,brown,nazar,cole,ormand,grigo,zhao,yeun,delion,singh,MAPLB,saxenaplb}, where the simultaneous emission of p-p subsystem of valence protons \cite{goldansky} is allowed after the tunneling through the barrier, but the sequential emission or 1p-decay is strongly suppressed by diproton correlations indicating the sensitivity of 2p-decay on pairing between the valence protons. The diproton weakly bound system in the decay channel might form a halo like structure \cite{xu1,LIN2009,xuxx2010} and support the enhanced probability of direct 2p-decay as shown by us \cite{saxenaplb} or may have deformed states \cite{wang2018} that are expected to impact the Q-values and consequently the decay modes of such diproton systems in competition with the $\beta$-decay and 1p-decay which needs investigation. \par

2p-decay life-time measurements have become a thrust area of research in recent times primarily due to the access to 2p-emitters, which has so far reached up to the light mass nuclei $^{45}$Fe, $^{54}$Zn, $^{48}$Ni, $^{19,20,22}$Mg, $^{30,31}$Ar, $^{22}$Si and $^{67}$Kr \cite{Ascher,mukha,Pftzner,Giovinazzo,Blank,Dossat,Pomorski2011,IMukhaar,Koldste,LIS,Wallace,LUND,SUN,xu}. Recent measurement of 2p-decay life-time of $^{67}$Kr has indicated the impact of shape deformation on 2p-radioactivity \cite{wang2018} and its sensitivity to the orbital configuration ($l$) and proton-proton interaction. 2p-decay life-time is sensitive to $l$ value as the centrifugal potential energy can reduce the tunneling probability and increase the half-life. The deformation dependency of 2p-radioactivity first speculated by Mukha \textit{et al.} \cite{mukha2006} and then theoretically investigated by us \cite{MAPRC89,MAPLB,saxenaplb,MAPRC90,MAPHYSCRTA} and more recently by Santhosh \cite{SanthoshCPPMDN} is expected to impact 2p-decay life-times, but it still remains a lesser explored area of research and is precisely the main objective of this work.\par

To explore the deformation effects on 2p-radioactivity, we first propose a semi-empirical relation to estimate 2p-decay half-lives by incorporating the quadrupole deformation in addition to the Q-value and angular momentum ($l$) sensitivity. The parameters of this relation are obtained in a way as to provide a best fit to the half-lives of experimentally known 2p-radioactive nuclei. As most of these proton rich nuclei lie close to the proton drip line, their measured values of deformations are not as yet available. Thus for our purpose we have instead utilized the theoretical values obtained from the relativistic mean-field (RMF) calculations \cite{geng,geng1,singh}. The formula thus obtained is found to perform much better than the other existing formulae \cite{Sreeja,Liu} and theories \cite{GLDM,Gonalves,Taveres,LiuGLM,ZouSEB,XingUFM,SanthoshCPPMDN,ZhuCPPM} for the 2p-decay life-times. Encouraged by this we have employed it to investigate and predict the possible candidates of 2p-emitters across the entire mass region with Z$<$82.\par

For many nuclei near the proton drip line even the accurate experimental Q-values are not available. Again we have taken recourse to using the theoretical RMF \cite{geng,geng1,singh} values for the 2p-decay energy Q$_{2p}$. However, 2p-decay half-lives being sensitive to the Q$_{2p}$, we have used the machine learning technique to minimize the uncertainties in theoretical Q$_{2p}$ values. Furthermore, 2p-emitters, being located near the drip line are expected to exhibit shape coexistence and possible halo formations. These two aspects have been demonstrated here by carrying out quadrupole constrained RMF calculations \cite{geng,geng1,singh} for the 2p-emitters which are experimentally known as well as those predicted by our formula.\par
\section{Formula for half-life}
The already existing semi-empirical formulae \cite{Sreeja,Liu} to evaluate the 2p-decay half-life consider the linear dependency of $log_{10}T_{1/2}$ on $Z_d^{0.8}/\sqrt{Q}$ in a similar way as done for $\alpha$-decay and 1p-decay. The $\alpha$-emission half-lives have shown significant improvement by including the deformation effects \cite{coban2012} which is also speculated for $\beta$-decay \cite{SARRIGUREN}, proton emission \cite{budaca2022} as well as two-proton emission \cite{budaca2022,SanthoshCPPMDN}. In fact, the proton and $\alpha$- emission both may be described by similar formulas with different parameters as both the decay modes are quantum tunneling effect. However, studies have shown \cite{dong2009} that the half-life is more sensitive to the Q-value for proton emission than that for the $\alpha$-decay, due to smaller reduced mass and also due to high centrifugal barrier. This implies that it is difficult to predict the half-life of proton emission for nuclei for which $l$ and Q-values are not available with good accuracy. Despite these difficulties, here we attempt to study the 2p-decay half-lives of nuclei across the mass region for Z$<$82 employing a semi-empirical relation which includes its explicit dependence on the Q-value, the reduced mass $\mu$, angular momentum $l$ carried by the emitted particles, and the quadrupole deformation $\beta$. This is expressed by
\begin{eqnarray}
log_{10} T_{1/2}&=& a+ b\sqrt{\mu}\sqrt{Z_d A^{1/3}}+c\sqrt{\mu}\left(\frac{Z_d}{\sqrt{Q}}\right)+ d\sqrt{l(l+1)}+e|\beta|^p
 \label{QF}
 \end{eqnarray}
where the symbols a, b, c, d, e, and p represent the parameters to be determined in a least square fit to the experimental data. Here, the half-life is in the unit of seconds, $Z_d$ is the proton number of the daughter nucleus, and $A$ denotes the mass number of the parent nucleus. The form of the first three terms is basically similar to that obtained in the semi-classical treatment for the evaluation of the expression for the transmission coefficient of an $\alpha$-particle through a Coulomb barrier \cite{gamow1928,segre1964}. The term $\sqrt{l(l+1)}$ reflects the hindrance effect of the centrifugal barrier \cite{Deng2021}. The form of the last term represents in a phenomenological way the effect of quadrupole deformation motivated by the detailed studies of Ref. \cite{froman1957} in which $\alpha$ emission probability has been found to increase by an exponential factor with increasing $\beta$. Since the nuclei being studied in this work lie at or beyond the proton drip line, the experimental values of $\beta$ for all these nuclei are not available as yet. Hence for our purpose in Eqn. (\ref{QF}),  we use the $\beta$ values obtained from the RMF theory \cite{geng,geng1,singh}. The $\beta$ deformation computed using the RMF approach are known to be in good agreement with those obtained using other theories, for example, Hartree-Fock-Bogoliubov (HFB) \cite{goriely}, Finite Range Droplet Model (FRDM) \cite{frdm2012}, and Weizsaecker-Skyrme-4 model (WS4) \cite{ws42014}. Also, so far there are no experimental data available for the 2p-radioactivity with $l\ne0$. However, as in Refs. \cite{Sreeja,Liu}, we have  used  2p-radioactivity half-lives of 7 nuclei with $l\ne0$ extracted from \cite{Gonalves} for our database, which are mentioned separately in Table \ref{table-comparison} wherein these have been indicated as 'other data'. The purpose of inclusion of such data for the fitting procedure is to test and also elucidate the  quality of results obtained with the present approach for the $l\ne0$ cases. Thus together we have used a set of 24 measured radioactive 2p-decay cases with experimental Q$_{2p}$ and log$_{10}$T$_{1/2}$ values as shown in Table \ref{table-comparison} for the fitting of Eqn. (\ref{QF}) as described briefly below.\par

\begin{figure}[!htbp]
\centering
\includegraphics[width=0.7\textwidth]{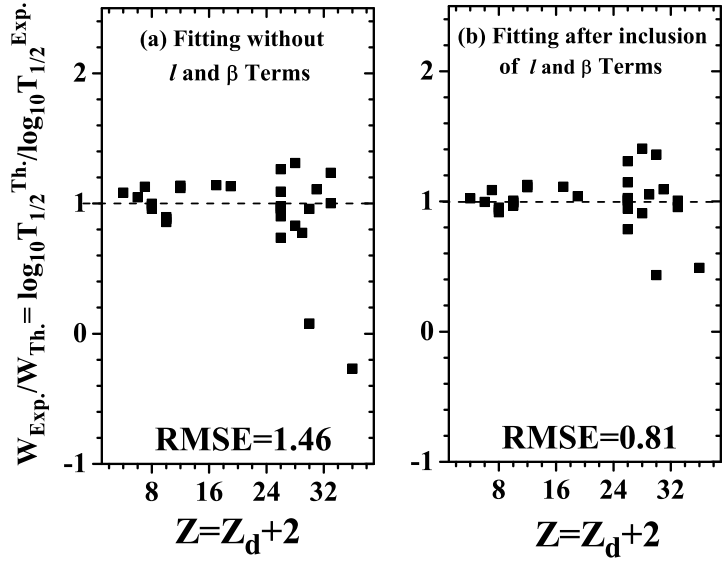}
\caption{Ratio of experimental to theoretical decay widths $W_{Exp.}/W_{Th.}=log_{10}T_{1/2}^{Th.}/ log_{10}T_{1/2}^{Exp.}$ for a best fit of our formula given by Eqn. (\ref{QF}) (a) without the angular momentum and deformation dependent terms, and (b) including the angular momentum and deformation dependent terms.}\label{rmse}
\end{figure}

\begin{landscape}
\begin{table*}[!htbp]
\caption{Comparison of present results (column 6) for the logarithmic half-lives log$_{10}$T$_{1/2}$ of 2p-emitters (column 1) with the experimental values (column 5) and other theoretical predictions (columns 7-16) \cite{Sreeja,Liu,GLDM,Gonalves,Taveres,LiuGLM,ZouSEB,XingUFM,SanthoshCPPMDN,ZhuCPPM}. For the experimental Q$_{2p}$, $l$ and half-life values the respective references are mentioned in the table. As the measured values for deformation $\beta$ are not available, we have used the values computed from the RMF theory (see text for details). Here abbreviations GLDM and ELDM denote the generalized and effective liquid drop models; GLM  stands for the Gamow- like model; SEB for the screened electrostatic barrier; UFM for the  unified fission model; CPPM for the Coulomb and proximity potential model whereas CPPMDN refers to the CPPM for the deformed nuclei.}
\centering
\resizebox{1.30\textwidth}{!}{%
\begin{tabular}{cccc|cccccccccccc}
\hline
\hline
\multicolumn{1}{c}{2p}&
\multicolumn{1}{c}{Q$_{2p}$}&
 \multicolumn{1}{c}{$l$}&
  \multicolumn{1}{c|}{$\beta$}&
  \multicolumn{12}{c}{log$_{10}$T$_{1/2}$ (sec.)}\\
   \cline{5-16}
Emitter& (MeV)&&&Data&Present&Sreeja & Liu &GLDM &ELDM &Taveres &GLM &SEB &UFM &CPPMDN &CPPM \\
&&&&&Formula&\cite{Sreeja}&\cite{Liu}&\cite{GLDM}&\cite{Gonalves}&\cite{Taveres}&\cite{LiuGLM}&\cite{ZouSEB}&\cite{XingUFM}& \cite{SanthoshCPPMDN}&\cite{ZhuCPPM}\\
  \hline
  \hline
&&&&Experimental Data&&&&&&&&&&&\\
$^{6}$Be &1.37$\pm$0.005 &0& 0.00&-20.30$^{+0.04}_{-0.02}$ \cite{Whaling}  &-21.04$\pm$0.09&-21.95&-23.81&-19.37$^{+0.01}_{-0.01}$&-19.97&                        &-19.70&-19.86&-19.41$^{+0.003}_{-0.003}$&-21.91&       \\ $^{12}$O &1.64$\pm$0.024 &0& 0.00&$>$-20.20 \cite{Jager}                   &-19.16$\pm$0.11&-18.47&-20.17&-19.17$^{+0.13}_{-0.08}$&-18.27&                        &-18.04&-17.70&-18.45$^{+0.04}_{-0.03}$  &-20.90&         \\
$^{12}$O &1.82$\pm$0.12  &0& 0.00&-20.94$^{+0.41}_{-0.19}$ \cite{Kekelis}  &-19.43$\pm$0.11&      &-20.52&-19.46$^{+0.13}_{-0.07}$&      &                        &-18.30&-18.03&-18.69$^{+0.15}_{-0.14}$  &-21.22&         \\
$^{12}$O &1.79$\pm$0.04  &0& 0.00&-20.10$^{+0.18}_{-0.13}$ \cite{Kryger}   &-19.39$\pm$0.11&      &-20.46&-19.43$^{+0.04}_{-0.03}$&      &                        &-18.26&-17.98&-18.65$^{+0.05}_{-0.05}$  &-21.17&         \\
$^{12}$O &1.80$\pm$0.40  &0& 0.00&-20.12$^{+0.78}_{-0.26}$ \cite{Suzuki}   &-19.40$\pm$0.11&      &-20.48&-19.44$^{+0.30}_{-0.20}$&      &                        &-18.27&-18.00&-18.66$^{+0.62}_{-0.41}$  &-21.19&         \\
$^{16}$Ne&1.33$\pm$0.80  &0& 0.44&-20.64$^{+0.30}_{-0.18}$ \cite{Kekelis}  &-20.23$\pm$0.22&-15.94&-17.53&-16.45$^{+0.23}_{-0.21}$&      &                        &-16.23&-15.47&-16.49$^{+0.24}_{-0.22}$  &-18.01&         \\
$^{16}$Ne&1.40$\pm$0.20  &0& 0.44&-20.38$^{+0.20}_{-0.13}$  \cite{Woodward}&-20.43$\pm$0.22&-16.16&-17.77&-16.63$^{+0.05}_{-0.05}$&-16.60&                        &-16.43&-15.71&-16.68$^{+0.05}_{-0.05}$  &-18.25&         \\
$^{19}$Mg&0.75$\pm$0.05  &0&-0.24&-11.40$^{+0.15}_{-0.35}$ \cite{Mukha2}   &-12.88$\pm$0.16&-10.66&-12.03&-11.79$^{+0.47}_{-0.42}$&-11.72&                        &-11.46&-10.92&-11.77$^{+0.47}_{-0.43}$  &-11.96&-12.17   \\
$^{45}$Fe&1.10$\pm$0.10  &0& 0.00&-2.40$^{+0.24}_{-0.15}$ \cite{Pftzner}   &-2.08 $\pm$0.20&      &-2.21 &-2.23$^{+1.34}_{-1.17}$ &      &                        &-2.09 &-2.30 &-1.94$^{+1.34}_{-1.18}$   &-2.76 &-2.07    \\
$^{45}$Fe&1.14$\pm$0.04  &0& 0.00&-2.07$^{+0.24}_{-0.21}$ \cite{Giovinazzo}&-2.57 $\pm$0.20&-1.66 &-2.64 &-2.71$^{+0.61}_{-0.57}$ &      &                        &-2.58 &-2.66 &-2.43$^{+0.61}_{-0.58}$   &-2.36 &-2.55    \\
$^{45}$Fe&1.21$\pm$0.05  &0& 0.00&-2.42$\pm$0.03  \cite{Audirac}           &-3.36 $\pm$0.20&-2.34 &-3.35 &-3.50$^{+0.56}_{-0.52}$ &      &                        &-3.37 &-3.27 &-3.23$^{+0.56}_{-0.52}$   &-3.15 &-3.33    \\
$^{45}$Fe&1.154$\pm$0.016&0& 0.00&-2.52$^{+0.12}_{-0.09}$ \cite{Dossat}    &-2.73 $\pm$0.20&-1.81 &-2.79 &-2.87$^{+0.19}_{-0.18}$ &-2.43 &                        &-2.74 &-2.79 &-2.60$^{+0.19}_{-0.18}$   &-2.53 &-2.71    \\
$^{48}$Ni&1.35$\pm$0.02  &0& 0.00&-2.08$^{+0.34}_{-0.52}$ \cite{Dossat}    &-3.10 $\pm$0.21&-2.13 &-3.13 &-3.24$^{+0.2}_{-0.2}$   &      &                        &-3.21 &-3.02 &-2.91$^{+0.21}_{-0.19}$   &-2.79 &-3.03    \\
$^{48}$Ni&1.29$\pm$0.04  &0& 0.00&-2.52$^{+0.41}_{-0.18}$ \cite{Pomorski}  &-2.48 $\pm$0.21&-1.61 &-2.59 &-2.62$^{+0.44}_{-0.42}$ &      &                        &-2.59 &-2.54 &-2.29$^{+0.44}_{-0.41}$   &-2.17 &-2.41    \\ $^{54}$Zn&1.48$\pm$0.02  &0& 0.27&-2.43$^{+0.26}_{-0.12}$ \cite{Blank}     &-3.48 $\pm$0.23&-1.83 &-2.81 &-2.95$^{+0.19}_{-0.19}$ &-2.52 &                        &-3.01 &-2.82 &-2.61$^{+0.19}_{-0.19}$   &-2.59 &-2.79    \\
$^{54}$Zn&1.28$\pm$0.21  &0& 0.27&-2.76$^{+0.19}_{-0.12}$ \cite{Ascher}    &-1.39 $\pm$0.24&-0.10 &-1.01 &-0.87$^{+0.25}_{-0.24}$ &      &                        &-0.93 &-1.24 &-0.52$^{+2.80}_{-2.18}$   &-1.45 &-0.71    \\
$^{67}$Kr&1.69$\pm$0.017 &0&-0.27&-1.70$\pm$0.18 \cite{goigoux}            &-0.99 $\pm$0.26&0.31  &-0.58 &-1.25$^{+0.16}_{-0.16}$ &-0.06 &                        &-0.76 &-0.87 &-0.54$^{+0.16}_{-0.16}$   &-1.06 &-0.22    \\
\hline
\multicolumn{4}{c}{}&
\multicolumn{1}{c}{'Other Data' \footnote{} \cite{Gonalves}}&
\multicolumn{11}{c}{}\\
\hline
$^{10}$N &1.30           &1& 0.25& -17.64                   &-18.77$\pm$0.14&-20.04&-18.59&                        &-17.64     &                        & -17.36 & -17.30 & -18.07                   &      &        \\
$^{28}$Cl&1.97           &2& 0.30&-12.95                   &-13.86$\pm$0.21&-14.52&-12.46&                        &-12.95     &-11.19$^{+0.31}_{-0.29}$& -13.11 & -12.16 &                          &      &         \\
$^{32}$K &2.08           &2&-0.12&-12.25                  &-12.16$\pm$0.19&-13.46&-11.55&                        &-12.25     &-10.62$^{+0.32}_{-0.30}$& -12.49 & -11.49 &                          &      &         \\
$^{52}$Cu&0.77           &4& 0.19&9.36                    &9.57  $\pm$0.28&8.62  &8.74  &                        &9.36       &                        & 8.94 &      &                          &      &         \\
$^{57}$Ga&2.05           &2& 0.25& -5.30                   &-5.18 $\pm$0.26&-5.22 &-4.14 &                        & -5.30     &-5.80$^{+0.51}_{-0.49}$ & -5.91 & -5.04 &                          &      &         \\
$^{60}$As&3.49           &4& 0.22& -8.68                   &-8.46 $\pm$0.27&-10.84&-8.33 &                        &-8.68      &-10.95$^{+0.42}_{-0.45}$& -9.40 & -8.10 &                          &      &         \\
$^{62}$As&0.69           &2& 0.22& 14.52                  &15.10 $\pm$0.29&13.83 &14.18 &                        &14.52      &                        & 14.06 &      &                          &      &         \\
\hline
\multicolumn{5}{c}{RMSE}&0.81&1.93&1.29&1.61&1.73&1.67&1.61&1.90&1.71&1.10&0.96\\
\hline
\hline \end{tabular}}
\label{table-comparison}
\end{table*}
\vspace{-2.0cm}
\tiny{\footnotetext{Since there are no experimental data available for the 2p-decay with $l\ne0$, following Refs. \cite{Sreeja,Liu} we have included the half-lives of 7 nuclei with $l\ne0$ extracted from Ref. \cite{Gonalves} in our database for the purpose of elucidation of the quality of results obtained using our present formula for the $l\ne0$ cases.}}
\end{landscape}

In our least square fit to the experimental data we obtain the root mean square error (RMSE) using the expression
\begin{eqnarray}\label{rmse-formula}
 RMSE &=& \sqrt{\frac{1}{N_{nucl}}\sum^{N_{nucl}}_{i=1}\left({x_i}\right)^2}
\end{eqnarray}

where $N_{nucl}$ is the total number of data used for the calculations. Here we use $x=(logT_{Th}-logT_{Exp})$, and $x = (Q_{Th}-Q_{Exp})$  for the RMSE evaluation of the half-lives and that of the Q-value, respectively. In order to elucidate the effect of deformation and centrifugal terms in Eqn. (\ref{QF}), first we have fitted our set of data by keeping only the first three terms. It yields the RMSE of 1.48 as also indicated in Fig. \ref{rmse} (a) where we have plotted the result of this fitting in the form of ratio of experimental to calculated decay widths $W_{Exp.}/W_{Th.}=log_{10}T_{1/2}^{Th.}/ log_{10}T_{1/2}^{Exp.}$ as a function of $Z=Z_{d}+2$. The obtained values of the coefficients a, b and c corresponding to this fitting are -22.3520$\pm$0.0646, -0.5323$\pm$0.0058 and 0.8667$\pm$0.0011, respectively. The RMSE obtained above reduces to a value of 1.14  when the fitting is done by including the fourth term related to the centrifugal potential. Thereafter, we apply the same procedure by including the last term  related to the quadrupole deformation $\beta$ in Eqn. (\ref{QF}). The power $p$ in the $\beta$ dependent term has been determined by varying its value and repeating the same fitting procedure to obtain a minimum of RMSE. The variation in the RMSE value has not been found significant for 2.5$\leq p \leq$3.5, and therefore, we have chosen $p=3$ for the formula. Finally, this procedure when all the five terms in Eqn. (\ref{QF}) are included yields a much reduced RMSE of 0.81. The results of the best fit have been shown in Fig. \ref{rmse} (b). It is gratifying  to note that the inclusion of deformation ($\beta$) dependence in Eqn. (\ref{QF}) reduces the RMSE significantly. Somewhat larger deviation for a few points in Fig. \ref{rmse} (b) for $Z>24$ suggests the need of more precise estimation from their particular experiments as our predictions are found in a good agreement for some other points of the same nuclei. The  values for the coefficients corresponding to the best fit are found to be  a$=$-21.3232$\pm$0.0719, b$=$-0.6491$\pm$0.0069, c$=$0.8703$\pm$0.0013, d$=$0.6909$\pm$0.0114, and e$=$-36.1448$\pm$1.1700, respectively. Evidently, the parameters a, b, d, and e are in units of sec. while c has the unit of sec.(MeV)$^{1/2}$. Clearly a positive value of the coefficient '$d$' indicates the hindrance effect of the centrifugal barrier, which can delay the transition thereby increasing the half-life. Contrary to this, the negative value of the coefficient '$e$' points out towards a greater transition probability for the deformed decaying nuclei as compared to the spherical case. \par

\begin{figure}[!htbp]
\centering
\includegraphics[width=0.8\textwidth]{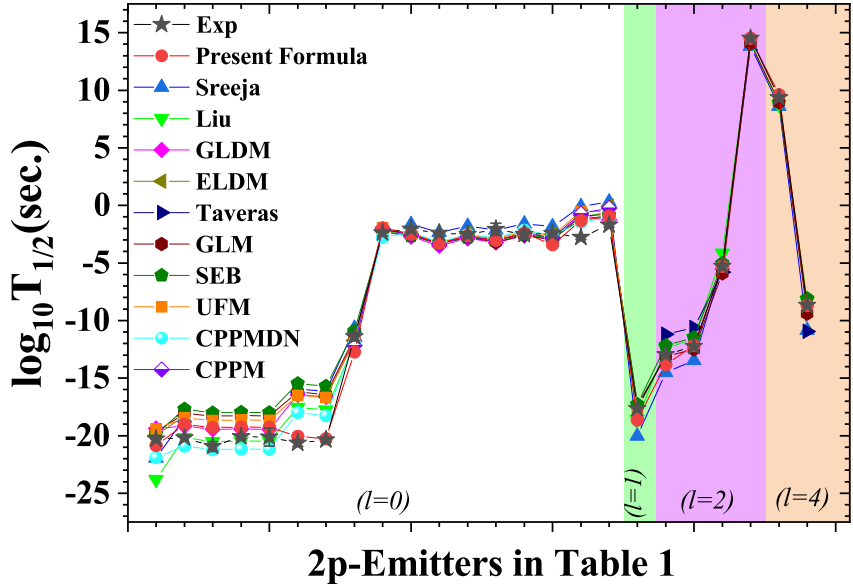}
\caption{(Colour online) Comparison of log$_{10}$T$_{1/2}$ values obtained using the present formula with the experimental data and the results obtained in other theoretical descriptions \cite{Sreeja,Liu,GLDM,Gonalves,Taveres,LiuGLM,ZouSEB,XingUFM,SanthoshCPPMDN,ZhuCPPM}. The results for $l\ne0$ have been separated from those with $l=0$ by the vertical lines in the figure. X-axis refers to the 2p-emitters of Table \ref{table-comparison} for which the half-lives have been displayed.}\label{half-life}
\end{figure}

Table \ref{table-comparison} shows the results for the 2p-decay half-life, in the form of log$_{10}$T$_{1/2}$, for the known 2p-emitters \cite{Whaling,Gonalves,Jager,Kekelis,Kryger,Suzuki,Woodward,Mukha2,Pftzner,Giovinazzo,goigoux,Audirac,Dossat,Pomorski,Blank,Ascher} calculated using our newly proposed formula given by Eqn. (\ref{QF}). It also lists the measured values of log$_{10}$T$_{1/2}$, Q$_{2p}$, and $l$ values used to set up the formula given by Eqn. (\ref{QF}). There are also given the deformation values $\beta$ used for these 2p-emitters, which in the absence of measurements are  taken to be those obtained from the RMF theory \cite{geng,geng1,singh} as described above. It is seen from the table that the calculated results are in excellent agreement with the experimental data. This agreement is especially found to be remarkable  for the deformed cases. Table \ref{table-comparison} also provides the calculated results for the half-lives along with the RMSE values obtained in other theoretical models for the purpose of comparison. It is readily seen that our proposed formula gives the least RMSE among all the available theoretical approaches \cite{Sreeja,Liu,GLDM,Gonalves,Taveres,LiuGLM,ZouSEB,XingUFM,SanthoshCPPMDN,ZhuCPPM}. A closer comparison of calculated results with experimental data in Table \ref{table-comparison} as well as in Fig. \ref{half-life} further shows the ability of present formula to provide a more satisfactory description of the experimental data as compared to those obtained from other theoretical approaches \cite{Sreeja,Liu,GLDM,Gonalves,Taveres,LiuGLM,ZouSEB,XingUFM,SanthoshCPPMDN,ZhuCPPM}. However, it is pointed out that the effect of deformation has very recently been taken into account by Santhosh \cite{SanthoshCPPMDN}  within the framework of Proximity Potential Model for the deformed nuclei. The calculated results for the half-lives in Ref. \cite{SanthoshCPPMDN} are seen to compare well with the experimental data, and comparatively are similar in quality to the predictions of the present approach.\par

\begin{table}[!htbp]
 \caption{2p-decay half-lives calculated by using present formula for new potential candidates as per experimental systematics. The Q$_{2p}$ values are taken from the latest evaluated nuclear properties table NUBASE2020 \cite{audi2020}. The angular momentum ($l$) is calculated from selection rules based on parity and spin of parent and daughter nuclei, which are also taken from NUBASE2020 \cite{audi2020}. As described in the text, deformation values $\beta$ have been taken from the RMF theory \cite{geng,geng1,singh}.}
 \centering
  \resizebox{0.6\textwidth}{!}{%
 \begin{tabular}{c@{\hskip 0.3in}c@{\hskip 0.3in}c@{\hskip 0.3in}c@{\hskip 0.3in}c}
 \hline
 \hline
 \multicolumn{1}{c}{2p}&
 \multicolumn{1}{c}{Q$_{2p}$ (MeV)}&
  \multicolumn{1}{c}{$l$}&
   \multicolumn{1}{c}{$\beta$}&
   \multicolumn{1}{c}{log$_{10}$T$_{1/2}$ (sec.)}\\
    Emitter&\cite{audi2020}&&&(Present Formula)\\
 \hline
 $^{22}$Si  &    1.58$\pm$0.50   &  0     &  0.00        &     -15.20$\pm$0.14  \\
 $^{39}$Ti  &    1.06$\pm$0.02   &  0     &  -0.15        &     -5.51$\pm$0.19   \\
 $^{42}$Cr  &    1.48$\pm$0.31   &  0     &  -0.17        &     -7.61$\pm$0.19   \\
 $^{49}$Ni  &    1.08$\pm$0.78   &  0     &  -0.06        &     0.06 $\pm$0.21   \\
 $^{55}$Zn  &    0.78$\pm$0.40   &  2     &  0.25         &     8.80 $\pm$0.27   \\
 $^{59}$Ge  &    1.60$\pm$0.45   &  0     &  0.21         &     -2.72$\pm$0.23   \\
 $^{64}$Se  &    0.70$\pm$0.52   &  0     &  0.24         &     14.26$\pm$0.27   \\
 $^{68}$Kr  &    1.46$\pm$0.54   &  0     &  -0.29        &     1.22 $\pm$0.27   \\
 $^{77}$Zr  &    0.44$\pm$0.46   &  0     &  0.48         &     32.87$\pm$0.42   \\
 $^{81}$Mo  &    0.73$\pm$0.58   &  0     &  0.57         &     16.96$\pm$0.50   \\
 $^{85}$Ru  &    1.13$\pm$0.64   &  0     &  -0.22        &     14.03$\pm$0.29   \\
 $^{104}$Te &    0.73$\pm$0.33   &  0     &  0.14         &     35.97$\pm$0.33   \\
 $^{108}$Xe &    1.01$\pm$0.39   &  0     &  0.17         &     27.28$\pm$0.32   \\
 $^{165}$Pt &    1.44$\pm$0.50   &  0     &  0.10         &     37.49$\pm$0.39   \\
 $^{170}$Hg &    1.85$\pm$0.34   &  0     &  0.00         &     29.87$\pm$0.38   \\
\hline
\hline \end{tabular}}
\label{half-life-expQ}
\end{table}
\section{Prediction of 2p-emitters}
The success of our theoretical approach described above encourages us to extend our study to look for the possible existence of other presently unknown 2p-emitters. For this purpose we have treated the following two cases separately.\\
(1) From amongst the proton rich nuclei across the mass region below Z$<$82 for which, the experimental data on Q$_{2p}$ and Q$_p$ are available in the latest database NUBASE2020 \cite{audi2020}, we identify the nuclei satisfying the criteria Q$_{2p}$ $>$ 0 and Q$_p$ $<$ 0 as the potential 2p-emitters and evaluate their half-lives using Eqn. (\ref{QF}). The results thus obtained for the logarithmic half-life have been listed in Table \ref{half-life-expQ}.\\
(2) The proton rich nuclei for which experimental data on Q$_{2p}$ and Q$_p$ values are not available as yet, we take recourse to using the theoretically computed results. In order to avoid the sensitivity and uncertainties of theoretical models we have used the following procedure. We have selected only those nuclei as possible potential 2p-emitter candidates for which, in addition to the RMF \cite{geng,geng1,singh} results for the Q$_{2p}$ and Q$_p$ values, other widely used theoretical models including NSM \cite{MAPLB,MAPRC89}, HFB \cite{goriely}, WS4 \cite{ws42014}, FRDM \cite{moller2019}, RCHB \cite{rchb2018}, KTUY05 \cite{ktuy2005}, and INM \cite{inm2012} fulfill the criteria of two-proton emission. This ensures that the results of the RMF theory, for  Q$_{2p}$, $l$ and $\beta$ we employ in Eqn. (\ref{QF}) to compute half-life, are consistent with those from other theoretical models.\par

\begin{figure}[!htbp]
\centering
\includegraphics[width=0.8\textwidth]{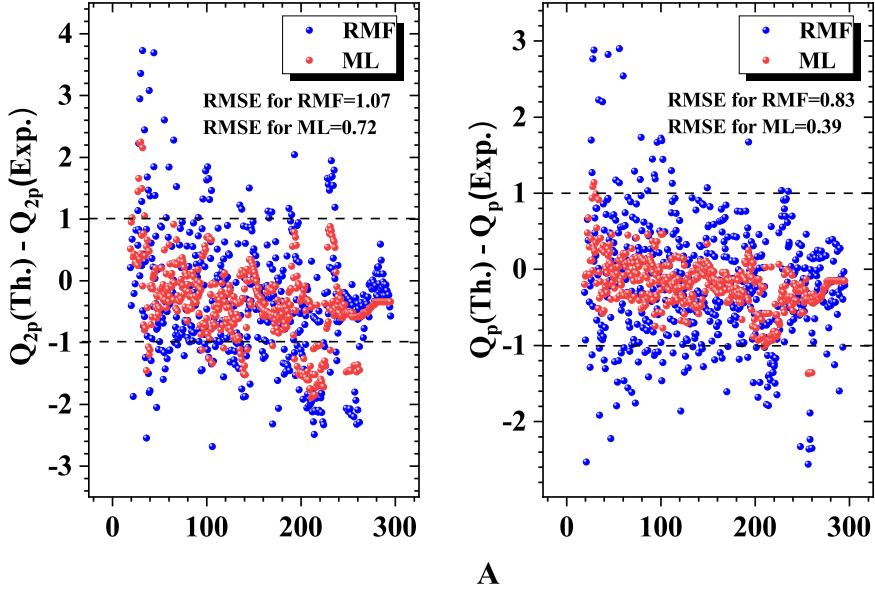}
\caption{(Colour online) Differences between theoretical and experimental Q$_{2p}$ and Q$_{p}$ values for the 529 and 531 nuclei, respectively. The blue dots denote the RMF error whereas that for the machine modified results are depicted using red dots. The RMS values for the errors corresponding to the RMF theory and that from the machine learning (ML) technique are also mentioned.}\label{ml}
\end{figure}

In recent times the use of machine learning techniques have been applied for the estimation of various data related to nuclear physics \cite{saxenajpg}.
These techniques can be utilized to bring down the possible theoretical uncertainties in our RMF computed Q$_{2p}$ values. With this in view, we apply one of the machine learning techniques by employing XGBoost algorithm, which was found to be effective in our earlier work on $\alpha$-emission \cite{saxenajpg}. For our purpose, we determine the possible errors in Q$_{2p}$ and Q$_p$ values from RMF theory using Q$_{2p}$ data for 2467 nuclei and Q$_p$ data for 2635 nuclei taken from the database NUBASE2020 \cite{audi2020} as the training data to machine learning. The estimation of errors from the machine algorithm has been found to qualify to a good accuracy which was validated by the test data of 529 and 531 nuclei, respectively, for Q$_{2p}$ and Q$_{p}$ values shown in Fig. \ref{ml}.
Hence the Q$_{2p}$ and Q$_p$ values from RMF theory for the potential nuclei are modified by including their possible errors obtained by the above-mentioned procedure. The modified Q$_{2p}$ and Q$_p$ values thus obtained are hereafter denoted by Q$_{2p}^{ML}$ and Q$_{p}^{ML}$, respectively. This set of Q$^{ML}$ values is now used to identify the possible new 2p-emitters by applying the criteria Q$_{2p}^{ML}$$>$0 and Q$_{p}^{ML}$$<$0. These have been listed in Table \ref{RMF-2pemitter} along with their calculated logarithmic half-life using Eqn. (\ref{QF}). Table \ref{RMF-2pemitter} also shows the Q$_{2p}$ and modified Q$_{2p}^{ML}$ values to illustrate how the machine learning technique affects the theoretically obtained RMF Q-values. In addition, the $l$ and $\beta$ values employed in the half-life calculations have also been given in the table. The results for the half-life show that apart from the nuclei $^{38}$Ti, $^{72}$Sr, $^{76}$Zr, and $^{84}$Ru, the other heavier possible candidates have relatively long life-time. For such nuclei in the heavy mass region, in addition to the 2p-decay, there always exist other possible dominant modes including $\alpha$, $\beta^+$, cluster decays as well as spontaneous fission.

\begin{table}[!htbp]
 \caption{Half-life of theoretical true 2p-emitter candidates obtained using the present formula where Q$_{2p}$ is taken equals the modified Q$_{2p}^{ML}$ (the disintegration energy which is modified by machine learning). The angular momentum ($l$) is calculated by using parity and spin of parent and daughter nuclei, which are taken from NUBASE2020 \cite{audi2020} and from Ref. \cite{mollerparity}. (see the text for detail).}
 \centering
  \resizebox{0.7\textwidth}{!}{%
 \begin{tabular}{c|c@{\hskip 0.3in}c@{\hskip 0.3in}|c|c|c}
 \hline
 \hline
 \multicolumn{1}{c|}{Potential}&
 \multicolumn{2}{c}{RMF}&
 \multicolumn{1}{|c}{Q$_{2p}^{ML}$}&
  \multicolumn{1}{|c|}{$l$}&
   \multicolumn{1}{c}{log$_{10}$T$_{1/2}$ (sec.)}\\
    \cline{2-3}
2p-Emitter&$\beta$&Q$_{2p}$ (MeV)&(MeV)&&(Present Formula)\\
   \hline
$^{38}$Ti    &  0.22      &  1.86  &   2.57  &   0      & -14.11$\pm$0.19   \\
$^{72}$Sr    &  -0.27     &  0.38  &   0.86  &   0      &  14.02$\pm$0.28    \\
$^{76}$Zr    &  -0.33     &  1.00  &   1.40  &   0      &  4.87 $\pm$0.30    \\
$^{84}$Ru    &  -0.22     &  0.60  &   1.20  &   0      &  12.63$\pm$0.28        \\
$^{105}$Te   &  0.12      &  0.24  &   0.58  &   2      &  46.37$\pm$0.37               \\
$^{120}$Nd   &  0.42      &  0.33  &   0.66  &   0      &  47.74$\pm$0.45           \\
$^{126}$Sm   &  0.40      &  0.24  &   1.07  &   0      &  31.48$\pm$0.42             \\
$^{150}$Hf   &  0.18      &  1.16  &   1.38  &   0      &  33.74$\pm$0.37               \\
$^{154}$W    & -0.11      &  1.25  &   1.65  &   0      &  29.25$\pm$0.37               \\
$^{159}$Os   &  -0.04     &  0.79  &   1.25  &   0      &  41.36$\pm$0.39            \\
\hline
\hline \end{tabular}}
\label{RMF-2pemitter}
\end{table}

\begin{figure}[h]
\centering
\includegraphics[width=0.6\textwidth]{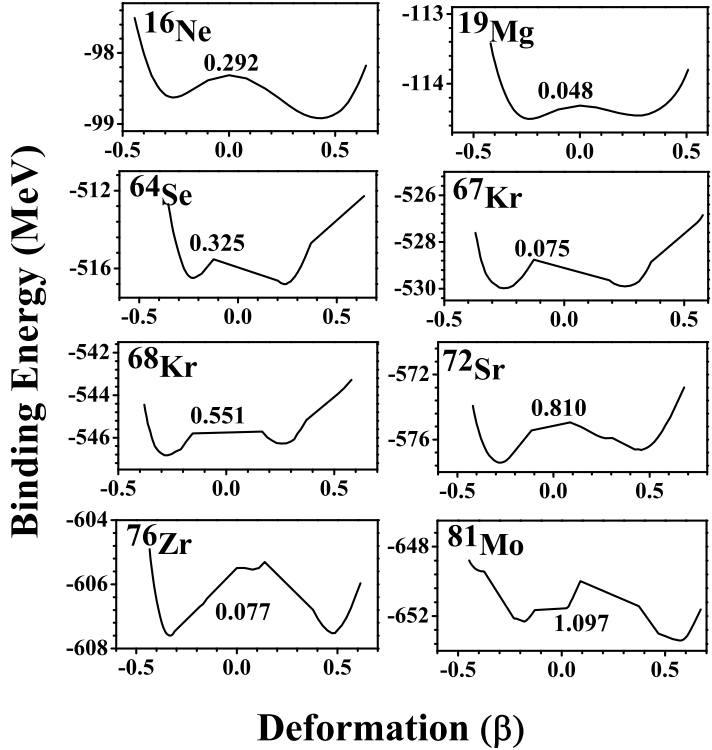}
\caption{Potential energy surfaces for 2p-emitters showing shape coexistence. The energy difference between prolate and oblate states are mentioned by numbers in the respective panels.}\label{fig2}
\end{figure}

\section{Shape-coexistence and Halos}
The neutron and proton rich nuclei in different regions of the nuclear chart are known to exhibit shape coexistence \cite{wood2016,wood1992}, wherein one or more states lying close to the ground state having different deformation ranging from oblate to spherical to prolate are found to exist. The shape coexistence in nuclei is known to provide a good evidence for the crucial role of interplay between the shell effects and the collective degrees of freedom in the structure and stability of nuclei. This has been amply demonstrated in several studies of nuclear decay and fission processes \cite{VM,nilsson1969,moller1994}. Thus in order to understand the influence of deformation on the process of 2p-radioactivity and the life-time of 2p-emitters, we have performed the quadrupole constrained RMF calculations \cite{geng,geng1,singh} for all the known 2p-emitters described in Table \ref{table-comparison}, and also of those predicted to be potential candidates as listed in Tables \ref{half-life-expQ} and \ref{RMF-2pemitter}. The results of our calculations show that out of 33 deformed cases considered altogether from the three tables, 17 nuclei exhibit shape coexistence with energy difference $\triangle$E between the two minima being small and having a range given by 40 keV$\leq$$\triangle$E$\leq$1 MeV. Typical potential energy surfaces obtained in such calculations for a few representative nuclei have been plotted in Fig. \ref{fig2} for the purpose of illustration. In several cases of the 2p-emitters shown in Fig. \ref{fig2} we observe two well separated  minima having prolate and oblate shapes. Strikingly the additional minima lying higher than the ground state have similar energy but very different shapes. Thus in such cases the nucleus may exist in a state corresponding  to the energy and shape of this additional close lying minimum. The life-time of this state will depend on the extent of overlap between its wave function with that of the ground state, its excitation energy and the height of saddle separating it from the ground state. A long life-time may give rise to a meta-stable state or even a shape isomer. We have also carried out the deformation constraint calculations for the daughter nuclei of the 2p-emitters. It is found that several of these daughter nuclei also have an additional minimum close to the ground state minimum but having different shapes as in the case of parent nuclei described above. Evidently, in a 2p-decay the transition from the parent to the daughter nuclei, apart from being strongly dependent on the Q$_{2p}$ energy, are expected to be influenced by their shapes. Thus 2p-decays involving different shapes for the parent and daughter nuclei are likely to be hindered resulting in longer half-lives. This in turn is expected to better facilitate the experimental probe of very short lived nuclei away from the drip lines. In view of the fact that the nuclei $^{16}$Ne, $^{19}$Mg, and $^{67}$Kr (shown in Fig. \ref{fig2}) along with $^{28}$Cl, $^{32}$K, and $^{60,62}$As being amongst the experimentally known 2p-emitters (see Table \ref{table-comparison}) exhibit shape coexistence, the other nuclei in our study showing shape coexistence, viz., $^{38,39}$Ti, $^{59}$Ge, $^{64}$Se, $^{68}$Kr, $^{72}$Sr, $^{76}$Zr, and $^{81}$Mo may be considered as possible 2p-emitters to be probed in the near future. It is observed that for these nuclei the half-lives obtained by using our present formula (Eqn. (\ref{QF})) as listed in Tables \ref{half-life-expQ} and \ref{RMF-2pemitter} fall in the range of values similar to those of the experimentally known 2p-emitters.\par

\begin{figure}[h]
\centering
\includegraphics[width=0.6\textwidth]{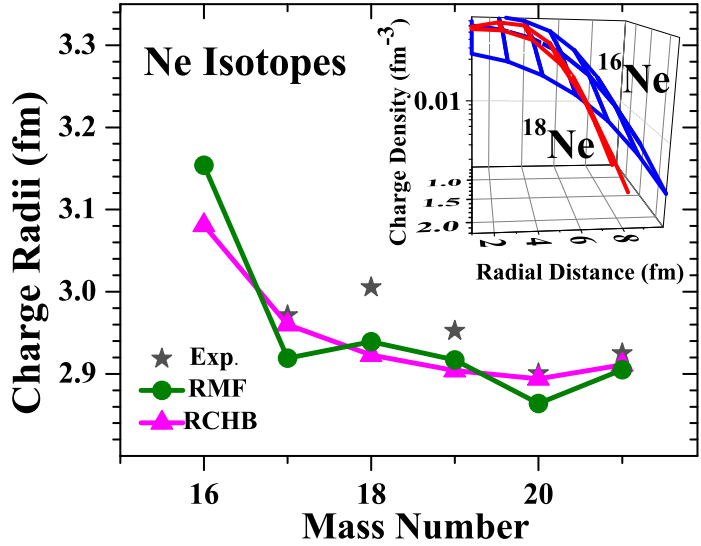}
\caption{(Colour Online) Calculated charge radii of Ne isotopes plotted as a function of mass number. One finds a sudden increase in the radius of $^{16}$Ne at the proton drip line indicating a halo formation. In the inset we compare the charge density distribution of $^{16}$Ne with that of the neighbouring isotope $^{18}$Ne. Again a much wider distribution for the $^{16}$Ne isotope is seen.}\label{fig3}
\end{figure}

As mentioned earlier, the ground states of the 2p-emitters studied here differ in shapes ranging from a deformed prolate or oblate to a spherical one. However, prolate shape in the ground state is found to be more prevalent. This predominance reaffirms the significance of prolate deformation in the 2p-radioactivity as observed by Mukha \textit{et al.} \cite{mukha2006}. Additionally, in the case of deformed nuclei, the occupancy in Nilsson orbital with the mixing of states near the Fermi level and the scattering of valence particles due to pairing correlations may contribute to a halo formation ~\cite{saxenajpg2021,yadav2004} with loosely bound pair of protons. With this in view, we look for the possible imprints of halo in the deformed nuclei listed in Tables \ref{table-comparison}, \ref{half-life-expQ}, \ref{RMF-2pemitter} by studying their charge radii and density profiles.
The charge radii $r_{c}=(\langle r^2_{c}\rangle)^{1/2}$ is calculated from the charge
density distributions using the following relation,
\begin{eqnarray}
         \langle r^2_{c}\rangle\,=\,\frac{\int \rho_{c}\, r^2
         d\tau}{\int \rho_{c}\,d\tau}
\end{eqnarray}
Our results show that the light nuclei having proton number Z$<$20, for example, $^{16}$Ne, $^{19}$Mg, $^{22}$Si etc. indeed display a sudden increase in their charge radii with respect to the neighbouring isotopes while approaching the proton drip line indicating thereby the halo like formation. The corresponding charge density profiles of such nuclei also show significant spatial spread as compared to the neighbouring isotopes. This has been illustrated for the case of $^{16}$Ne in Fig. \ref{fig3} which shows the charge radii of Ne isotopes as a function of  mass number. A sudden increase in the charge radius for the proton drip line isotope $^{16}$Ne is seen. The inset in Fig. \ref{fig3} shows the density profile of $^{16}$Ne in comparison with that of its  neighbouring isotope $^{18}$Ne to highlight the spread of charge density in the case of isotope $^{16}$Ne with a halo. Here the occurrence of halo formation is found only for the lighter nuclei with Z$<$20 which is in accord with our earlier prediction of 2p-halo in the light mass region \cite{saxenaplb}.

\section{Conclusion}
Deformation dependence of the two-proton (2p) emission half-lives for nuclei has been studied within the framework of a semi-empirical formula akin to that used for the alpha radioactivity treated as quantum tunnelling phenomenon. Apart from the usual Q-value and angular momentum dependence, it incorporates the quadrupole deformation in a phenomenological way. The method is found to be robust and performs quite well in comparison to other descriptions \cite{Sreeja,Liu,GLDM,Gonalves,Taveres,LiuGLM,ZouSEB,XingUFM,SanthoshCPPMDN,ZhuCPPM} for the 2p-radioactivity half-lives. The results for the identified potential 2p-emitters have been listed in Tables \ref{half-life-expQ} and \ref{RMF-2pemitter}. It is expected that some of these predicted 2p-radioactive nuclei would be accessible in the near future for measurements and possible verification providing a new impetus to the field of radioactivity.\par
Furthermore, we have shown that many of the 2p-emitters being located close to the proton drip-line exhibit shape coexistence with a small energy difference between the oblate and prolate shapes. However, for the ground state the prolate shape is found to be more prevalent. In addition, it is found that the lighter 2p-emitters with Z$<$20 being excessively proton rich (Z/N$>$1.4) tend to have halo formation due to the loosely bound protons.
\section{Acknowledgements}
Authors would like to thank M. Kaushik and Prafulla Saxena for their help for the calculations using machine learning techniques. Authors G. Saxena and M. Aggarwal gratefully acknowledge the support provided by Science and Engineering Research Board (SERB)-DST, Govt. of India under CRG/2019/001851 and WOS-A scheme, respectively.

\end{document}